# Probing the topological charge of a vortex beam with dynamic angular double slits

Dongzhi Fu, Dongxu Chen, Ruifeng Liu, Yunlong Wang, Hong Gao, Fuli Li, and Pei Zhang *

*MOE Key Laboratory for Nonequilibrium Synthesis and Modulation of Condensed Matter, Department of Applied Physics, Xi'an Jiaotong University, Xi'an 710049, China*
*Corresponding author: zhangpei@mail.ustc.edu.cn*

When a vortex beam with the spiral phase structure passes through a dynamic angular double slits (ADS), the interference pattern changes alternatively between destructive and constructive at the angular bisector direction of the ADS due to their phase difference. Based on this property, we experimentally demonstrate a simple method, which can precisely and efficiently determine the topological charge of vortex beams. What's more, this scheme allows determining both the modulus and sign of the topological charge of vortex beams simultaneously.

As is well known, photons carry both spin and orbital angular momentum (OAM) [1-3]. The photon spin is associated with the polarization and the OAM is associated with the azimuthal phase $\exp(il\varphi)$, where $\varphi$ is the azimuthal angle and $l$ is the azimuthal index referring to the topological charge (TC) of optical vortex [3]. Such beam has a singularity on the propagating axis, resulting in a dark point in the center of the transverse intensity distribution and carrying an OAM of $l\hbar$ per photon [3-6]. As $l$ can take any integer value, it offers an unbounded state space for the quantum information processing [7-10]. In the last decades, the interest for photon OAM spreads over a lot of fields, such as quantum computation [11-14], quantum cryptography [15], high capacity optical communication [16-18], optical microscopy [19] and micromanipulation [20, 21], thus characterizing the TC of vortex beams becomes more important.

Several methods have been proposed to determining the OAM states. Observing interference pattern is a common way to analyze the phase structure of vortex beams, such as interference with a plane wave [22, 23] or with its mirror image [24], double-slit interference [25], multiple-pinhole interference [26], triangular aperture diffraction [27], annular aperture diffraction [28] and Cartesian to log-polar coordinate transformation [29-31]. Some other interference methods have also been proposed for the same purpose, for example, Padgett's group proposed a scheme to efficiently sort different OAM states under the single photon level based on a Mach-Zehnder interferometer with Dove prisms inserted into each arm [32, 33]. Recently, there is a refractive beam copying method which can efficiently sort OAM states by a complex optical transformation [34, 35]. For these methods, most of them are involved with a complexity of interferential patterns or a complicated experimental setup.

In our previous study [36], we found that the interference pattern of the vortex beam after passing through an angular double slits (ADS) changes alternatively between darkness and brightness at the center of the ADS because of the phase difference. Moreover, it has opposite variation tendencies between the transition of constructive and destructive interference for opposite signs of the TC of vortex beams. So the modulus and sign of the TC of the vortex beam can be determined according to the periodic and moving direction of interference patterns. However, this scheme needs to collect a series of interference patterns and estimate the periodic of interference patterns, which is not precise. Based on this work, we describe a method by which the modulus and sign of the TC of vortex beams can be conveniently, precisely and simultaneously determined.

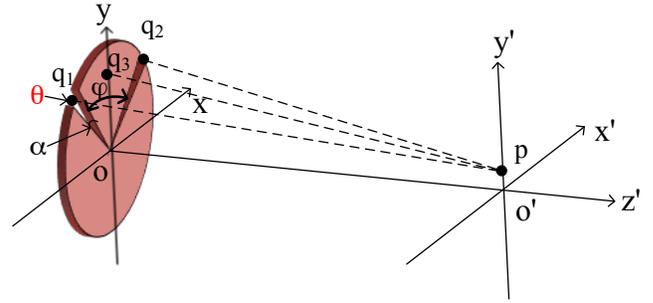

Fig. 1. Schematic of an ADS interference. $\alpha$ is the width of an angular-single-slit, and $\varphi$ is the angle between dynamic ADS. $q_1$, $q_2$ and $q_3$ are three points on the mask and $oq_3$ is the angular bisector of $\angle q_1 o q_2$. $p$ is a point on the far-field which is chosen to observe interferential intensity. $\theta$ represents an additional phase on one single slit.

Figure 1 shows the schematic of the scheme with an ADS mask. When a vortex beam with a spiral phase front $\exp(il\varphi)$ passes through dynamic ADS along the $z'$-axis, a constructive or destructive interference pattern occurs at the $o'p$-axis according to the optical path difference between $q_1 p$ and $q_2 p$. The phase difference between $q_1 p$ and $q_2 p$ is

$$\Delta\phi = l\varphi + 2\pi \frac{|q_1 p| - |q_2 p|}{\lambda}, \qquad (1)$$

Where $\lambda$ is the wavelength of the illuminating light. Then the intensity at $p$ is

$$I = |E_1 + E_2|^2 \propto 2\cos^2\frac{\Delta\phi}{2} = (1 + \cos\Delta\phi), \qquad (2)$$

where $E_1$ and $E_2$ are the electric field from the two single-slits. If dynamic ADS is symmetric with respect to the $o'p$-axis, the optical path difference between $q_1p$ and $q_2p$ is zero, and the phase difference between $q_1p$ and $q_2p$ is only dependent on the transverse phase with the form of $\Delta\phi = l\varphi$.

If $\Delta\phi = N\pi$ (where $N$ is an integer), a constructive interference pattern occurs at the point $p$ when $N$ is even, and a destructive interference pattern occurs at the point $p$ when $N$ is odd. For a measured vortex beam, we will get constructive or destructive interference patterns on the $o'p$-axis due to the spiral phase. As shown in Fig. 1, the angular bisector direction of dynamic ADS is parallel to the $y$-axis and $y'$-axis. We can fix the angular bisector at $y$-axis, and continuously rotate the two single-slits with the same angle with respect to the $y$-axis simultaneously. Thus, a periodic constructive or destructive interference pattern can be obtained at the $y'$-axis and the TC of the vortex beam is equal to the number of intensity period when $\varphi$ varying from 0 to $2\pi$. In order to determine both the modulus and sign of the TC of vortex beams simultaneously, we can add an additional phase $\theta$ to one of the ADS, and then the phase difference between $q_1p$ and $q_2p$ turns into.

$$\Delta\phi = l\varphi + \theta. \quad (3)$$

In this case, the additional phase leads to a displacement of intensity period (or a rotation in polar coordinates of the $\varphi - I$ curves as we show below), and the direction will indicates the sign of the TC. Based on this property, we propose a simple, quick and accurate method for measuring both the modulus and sign of the TC of vortex beams simultaneously.

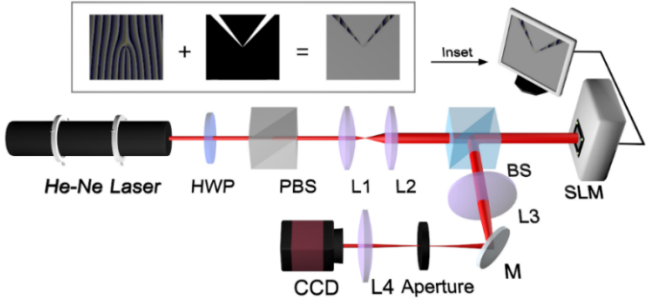

Fig. 2. A sketch of the experimental setup. The inset shows the mask pattern loaded on the SLM.

The experimental setup of the ADS interference is shown in Fig. 2. Light from a He-Ne laser passes through a half-wave plate (HWP) and a polarizing beam splitter (PBS), which are used to adjust intensity and filter the polarization. Before vertically illuminating on a spatial light modulator (SLM), it is expanded with two lenses, L1 and L2. SLM is used to generate vortex beams and ADS. The inset of Fig. 2 shows the mask, which consists of a fork grating and ADS (in the experimental process, the angular-single-slit width is $8^o$). The light of first-order diffraction is selected by an aperture, and then recorded by a charge-coupled device (CCD) camera. For the simplicity, we fix the angular bisector on the $y$-axis, so the center of interference patterns will not move when the angle of ADS is changed. By integrating the intensity in the center, we can get the relation between intensity and angle of ADS, which will reveal the information of the TC.

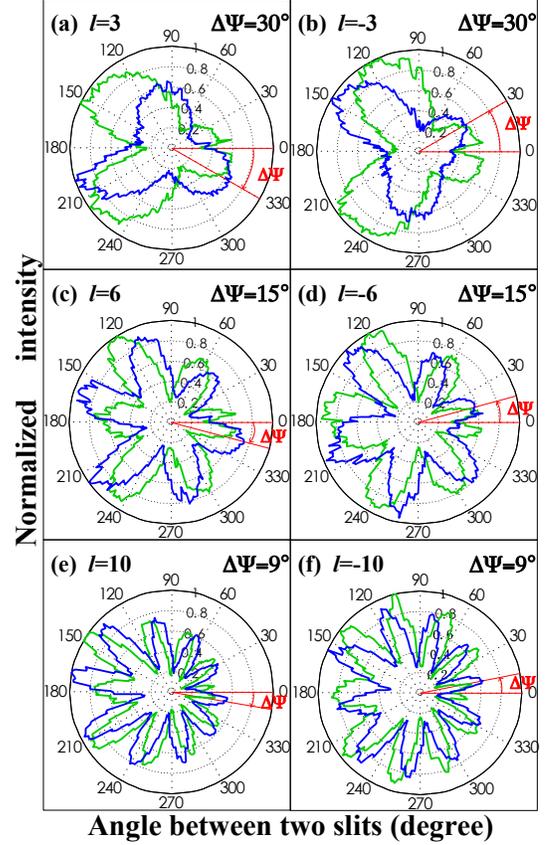

Fig. 3. The variation of intensity with different OAM states and the rotation of the $\varphi - I$ curves in polar coordinates when the additional phase $\theta$ equals $\pi/2$. The first column shows that the sign of the TC is positive and the other is negative. The green curves are the $\varphi - I$ curves without additional phase in the ADS and the blue curves are with an additional $\pi/2$ phase. $I$ is the intensity of interference patterns at the angular bisector direction of ADS. $\varphi$ is the angle between the two slits. $\Delta\Psi$ is the rotation angle value of the $\varphi - I$ curve.

Figure 3 shows experimental results of determining the TCs of different LG beams with dynamic ADS. The green curves are gotten by setting dynamic ADS without additional phase, thus they can only reveal the modular values of TC from the period of intensity. The blue curves show the rotation due to the additional phase in the ADS. By comparing the blue curves with the green curves, we find that when the sign of the TC is positive, the rotating direction of the $\varphi - I$ curves is clockwise. In contrast, when the sign of the TC is negative, the direction of rotation is anticlockwise. So with the help of an additional phase, we can both determine the modulus and sign of the TC of vortex beams simultaneously. According to the Eqs. (2) and (3), the interference intensity can be written as

$$I \propto \{1 + \cos[l(\varphi + \frac{\theta}{l})]\}. \quad (4)$$

which shows that the $\varphi - I$ curves will rotate $\Delta\Psi = \theta/l$ degrees in polar coordinates. In our experiment, we choose the additional phase $\theta = \pi/2$, thus the $\varphi - I$ curves rotate $\pi/(2l)$ clockwise (anticlockwise) when $l$ is positive

(negative). Because of the periodicity of the intensity distribution, there is degeneracy for opposite signs of OAM states when the additional phase $\theta$ equals $N\pi$. So the additional phase $\theta = \pi/2$ will get a better distinguishability of the TC sign than other angles. Furthermore, there is an interesting phenomenon that a split peak appears near $2N\pi$. This can be explained by considering both the interference and the transmitted light intensity. The interference peak appears at $2N\pi$, while the total transmitted light intensity of the ADS decreases around $2N\pi$ due to the width of angular-single-slit and the overlap of two slits around $2N\pi$. So the superimposition of these two mechanisms leads to the peak splitting. This is sketched in Fig. 4(a), where the black curve (split peak) is resulted from the superimposition of the blue curve (interference pattern) and the red curve (total transmitted intensity). When we add an additional phase $\Delta\Psi$ to one of the ADS, the blue solid line is offset and the movement direction is decided by the sign of the TC, as can be seen from Fig. 4(b), but the red solid line is the same as before, so the $\varphi - I$ curves which is expressed as the black solid line is changed, especially around the $2m\pi$, where appears a split peak. However, the split peak is hard to be observed in experiment because the power of one of the peaks is too low. We experimentally test $l = \pm 3, \pm 6, \pm 10$, the results clearly present the modulus and sign of the TC of vortex beams and have a good agreement with the theory.

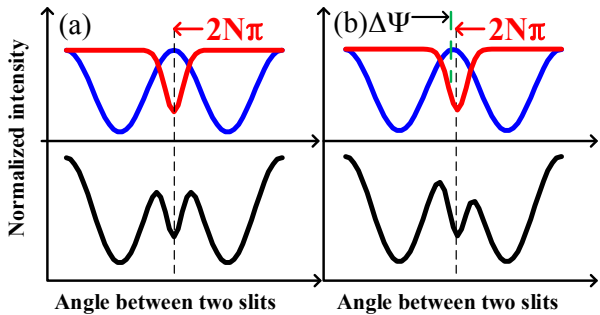

Fig. 4. Explanation of the peak split. The blue and red solid lines denote the interference intensity distribution and the intensity of light transmitted ADS, respectively. The black solid line represents the $\varphi - I$ curve in Cartesian. The black dashed line denotes the location $\varphi = 2N\pi$, where $N$ is an integer. The green dashed line denotes the offset of the blue solid line when we add an additional phase $\Delta\Psi$ to one of the ADS.

The $\varphi - I$ curves are not strictly symmetry in Fig. 3 because of the width of angular-single-slit and the impact of noise. However, determining the modulus and sign of light's TC is unaffected. What's more, the additional phase is not hard to be implemented without SLM in the application. For example, a glass plate can be inserted in one slit of the ADS to achieve a fixed phase. And also, a PD instead of CCD can be used to record the central intensity of interference patterns, which could make the detection faster (meantime, a lens should be inserted before the PD to collect the intensity of interference pattern) [37]. So probing the TC of vortex beam with dynamic ADS is practical under current technology.

In conclusion, we propose an interference method for probing the TC of a vortex beam with dynamic ADS and demonstrate experimentally that this method can be used to measure both the modulus and sign of a vortex beam's TC simultaneously and conveniently. When a vortex beam with a spiral phase structure illuminates the ADS, the far-field interference patterns depend on the phase difference between the two slits. So we can obtain periodic $\varphi - I$ curves by recording the intensity at the angular bisector direction of the ADS when scanning the angle between the two slits. By adding an additional phase in one angular slit, we can precisely determine the modulus and sign of light's TC simultaneously. What's more, this experiment clearly reveals the spiral phase structure of vortex beams, and our method without complicated experimental setup or profound results allows us to better understand a vortex beam. Our scheme holds promise to detect different OAM modes with a very simple and low-cost structure.

The authors thank Jian-ji Dong and Hai-long Zhou for helpful discussions. This work is supported by the Fundamental Research Funds for the Central Universities and the National Natural Science Foundation of China ( Grant Nos. 11374008, 11374238 and 11374239).